\documentclass[10pt,amssymb]{article}

\usepackage[english]{babel}
\usepackage{fullpage}
\usepackage{here}
\usepackage{amsmath}
\usepackage{amsfonts}
\usepackage{hyperref}
\usepackage{graphicx}

\def\ahad{a_\mu(\mathrm{had},\mathrm{LO})}

\begin{document}

\title{Dispersive estimation of the LO hadronic contribution to the muon ${g-2}$: fitting incompatible ${e^+e^-}$ data 
}

\author{V.\ V.\ Bryzgalov\footnote{Valery.Bryzgalov@ihep.ru}\, and\,
        O.\ V.\ Zenin\footnote{Oleg.Zenin@ihep.ru}\\
                {\it NRC ``Kurchatov Institute'' -- IHEP, Protvino, Russia}}

\date{}
\maketitle

\begin{abstract}
Using an up-to-date compilation of $\sigma_{\mathrm{tot}}(e^+e^- \to hadrons)$ data
we estimated the LO hadronic contribution to the muon anomalous magnetic moment, $\ahad$.
Incompatibilities between $\sigma_{\mathrm{tot}}(e^+e^- \to hadrons)$ 
measurements by independent experiments are mitigated by extra systematic uncertainties 
estimated using as a guideline a requirement of uniformity of $\chi^2$ distribution over degrees of freedom in joint fits of $\sigma_{\mathrm{tot}}$.
Tensions in the $e^+e^-$ input data translate into an expanded uncertainty of the 
$\ahad = (697.7 \pm {9.8}_{e^+e^-} \pm 3.6_{sys}) \times 10^{-10}$.
Given this, we obtain the SM prediction for the muon anomaly 
$a_\mu^{\mathrm{SM}} =  11659185.5(10.6) \times 10^{-10}$,
below the experimental world average $a_\mu^{\mathrm{exp}}$ at {$2\sigma$} level.
\end{abstract}

\section{Introduction} \label{sec:intro}

\begin{figure}[H]
	\begin{center}
	\includegraphics[width=0.2\textwidth,clip]{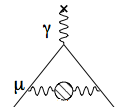}
	\end{center}
\caption{
	The leading order hadronic contribution to the $a_\mu$. 
	The blob represents 1-particle irreducible VP operator induced by hadronic EM currents. 
} \label{fig:hadlo}
\end{figure}

The anomalous magnetic moment of the muon $a_\mu$ is the most precisely measured quantity in particle physics \cite{Muong-2:2026qnz}
sensitive to physics beyond the SM \cite{Aliberti:2025beg}.
The accuracy of $a_\mu$ calculation in the SM  is limited by the leading order hadronic contribution 
to the photon vacuum polarization (VP) operator in the triangle diagram (Fig.~\ref{fig:hadlo}).
As the amplitude is dominated by momenta running through the photon line in the loop $|Q^2| \sim m_\mu^2 < 1$~GeV$^2$, 
it cannot be calculated in the perturbative QCD. 
Despite a remarkable progress in precision of the lattice QCD calculations \cite{Aliberti:2025beg}, 
it is desirable to have an independent method of the amplitude evaluation.
Currently, the dispersive calculation of the photon VP based on measurements 
of the total hadronic cross section with $e^+e^-$ beams \cite{Petermann:1957ir} 
\footnote{Optionally supplemented by hadronic form-factors from $\tau$ lepton decays.}
is the only viable alternative.%

The total cross section $\sigma_{\mathrm{tot}}(e^+e^-  \to hadrons)$ is measured 
in exclusive final states ($\pi^+\pi^-$, $\pi^+\pi^-\pi^0$, $\pi^0\gamma$, $K\bar{K}$, {\it etc.})
at $\sqrt{s} < 2$~GeV and inclusively ($\ge$ 2 hadrons) at $\sqrt{s} > 2$~GeV. 

\begin{figure}[H]
	\begin{center}
		\includegraphics[width=0.8\textwidth,clip]{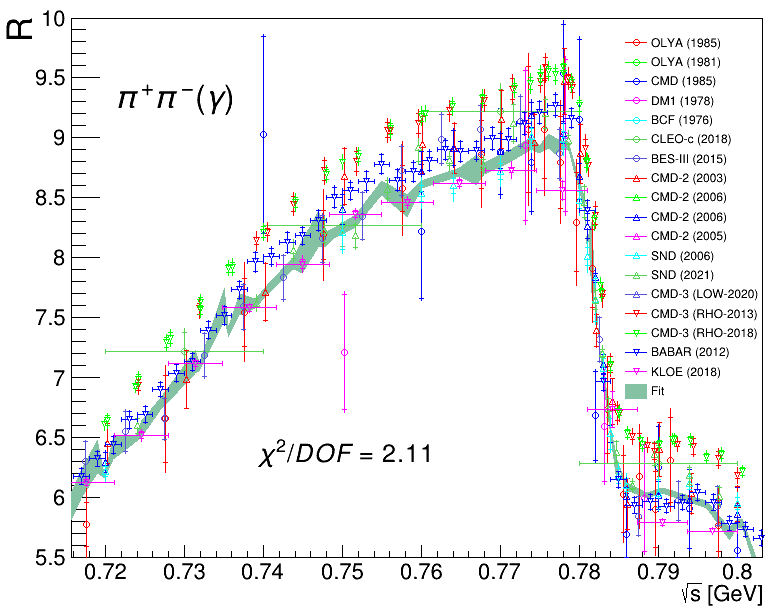}%
	\end{center}
	\caption{ $R^{\pi^+\pi^-(\gamma)}$ in the $\rho$--$\omega$ interference region, 
	most representative to demonstrate tensions between  BaBar, CMD-3 and KLOE $\sigma_{\mathrm{tot}}(e^+e^- \to \pi^+\pi^-(\gamma))$ measurements.
	Total experimental uncertainties are shown by vertical error bars with ticks indicating the statistical uncertainty.
	The fit is performed in the range $0.3 < \sqrt{s} < 2.0$~GeV 
	with unmodified experimental uncertainties.
	The fit uncertainty (shown by green band) is scaled by $\sqrt{\chi^2/n_{\mathrm{dof}}}$.
} \label{fig:2pi}
\end{figure}

The leading contribution ($\simeq 70\%$) to the dispersion integral \cite{Petermann:1957ir}, 
\begin{eqnarray} 
	a_\mu({\mathrm{had, LO}}) &=& 
		4\alpha_0^2 \int^{\infty}_{m_\pi^2} 
		\frac{ds}{s} K(s)\, \frac{1}{\pi}\, \mathrm{Im}\, \Pi^{\mathrm{had}}(s) = 
		\frac{\alpha_0^2}{3\pi^2} \int^{\infty}_{m_\pi^2} 
		\frac{ds}{s} K(s) R^{\mathrm{had}} (s)  
	 \label{eq:dispersion} \\
	R^{\mathrm{had}}(s) &=&
		\sigma_{\mathrm{tot}} (e^+e^- \to \gamma^* \to hadrons,\, \mathrm{bare}\,\, e^+e^-\,\, \mathrm{vertex\,\, and}\,\, \gamma^*) \,
		\left/  \,
		\frac{4\pi\alpha_0^2}{3s} 
		\right. 
	  \label{eq:R} \\
	K(s) &=& \int^1_0 dx \frac{x^2 (1-x)}{x^2 + (1-x) (s/m_\mu^2)}  \, , \nonumber
\end{eqnarray}
comes from $\sigma_{\mathrm{tot}}(e^+e^- \to \pi^+\pi^-(\gamma))$ at $2m_\pi < \sqrt{s} < 2$~GeV, 
where the cross section was measured with $\sim 1\%$ precision by 
BaBar \cite{BaBar:2012bdw}, 
KLOE-2 \cite{KLOE-2:2017fda}
and CMD-3 \cite{CMD-3:2023alj} experiments.
However, despite the claimed precision, these measurements are mutually incompatible at $\sim 3-5\sigma$ level (Fig.~\ref{fig:2pi}).
Possible sources of tensions are widely discussed in the literature, to no definite conclusion so far~\cite{Aliberti:2025beg}.
Being agnostic about instrumental origin of the tensions, 
we assume that incompatible measurements are affected by systematic effects 
that might not be controlled by the experiments {\it in situ},
but still can be estimated by comparing their published results.
The estimate of the uncontrolled systematic uncertainty facilitates more consistent 
fits of $\sigma_{\mathrm{tot}}(e^+e^- \to hadrons)$ in all final states 
and, consequently, the dispersive evaluation of $\ahad$ with a realistic uncertainty.

\section{Fitting (incompatible) $\sigma_{\mathrm{tot}}(e^+e^- \to hadrons)$ data}

Evaluation of the dispersion integral (\ref{eq:dispersion}) involves 
rescaling of published $\sigma_{\mathrm{tot}}(e^+e^- \to hadrons)$ measurements to the $R$-ratio (\ref{eq:R}),
a sufficiently smooth parameterization of the latter for each hadronic final state (channel, in what follows), 
and fitting the parameterized  $R^{\mathrm{had}}(s)$ by minimization of the $\chi^2$:
\begin{equation}\label{eq:chi2}
	\chi^2 = \sum\limits_{i,\, k} \sum\limits_{j,\, l} %
		   \left[ \frac{1}{(\Delta \sqrt{s_{ik}})} \int\limits_{(\Delta \sqrt{s_{ik}})} R^{\mathrm{had}}_{fit}(s) d\sqrt{s} - R^{\mathrm{had}}_{ik} \right] %
		   \times \left(C^{-1}\right)_{ik\, jl} \times %
		   \left[ \frac{1}{(\Delta \sqrt{s_{jl}})} \int\limits_{(\Delta \sqrt{s_{jl}})} R^{\mathrm{had}}_{fit}(s) d\sqrt{s} - R^{\mathrm{had}}_{jl} \right] \,  .%
\end{equation}
Here and in what follows indices $i$, $j$ run over experiments and $k$, $l$ over their $s$ bins.
$R^{\mathrm{had}}_{fit} (s)$ is the fitted parameterization, 
$R^{\mathrm{had}}_{ik}$ are the measurements in $s_{ik}$ bins, 
and $C$ is the covariance matrix parameterized as
\begin{eqnarray} \label{eq:C}
	C_{ik\, jl} = \delta_{ij}\delta_{kl} \sigma^2_{\mathrm{stat}, ik}  & + &  %
						  \frac{1}{(\Delta\sqrt{s_{ik}})} \int\limits_{(\Delta\sqrt{s_{ik}})} R^{\mathrm{had}}_{fit}(s) d\sqrt{s} \times 
						  \frac{1}{(\Delta\sqrt{s_{jl}})} \int\limits_{(\Delta\sqrt{s_{jl}})} R^{\mathrm{had}}_{fit}(s) d\sqrt{s} \, \times  \nonumber \\
						  &&  \times \Delta_{\mathrm{sys}, ik}  \Delta_{\mathrm{sys}, jl} \times
						      \left( \delta_{ij} + c_{ij} \right)\, ,
\end{eqnarray}
where $\sigma_{\mathrm{stat}, ik}$ and $\Delta_{\mathrm{sys}, ik}$ are, respectively,
statistical and relative systematic uncertainties in the $k$-th bin of the $i$-th experiment,
and $c_{ij} = c_{ji}$ ($c_{ii} = 0$) are optional correlations between the experiments.
The numerical procedures are described in more detail elsewhere \cite{Bryzgalov:2024ebj}. 
The program code and an up-to-date index of the $e^+e^- \to hadrons$ input data are available online \cite{the-code,ihep-cs}. 

The $\chi^2$ of a consistent fit lies with 95\% probability  in an approximate interval  
$n_{\mathrm{dof}} \pm 2\sqrt{2n_{\mathrm{dof}}}$,
for the number of degrees of freedom 
$n_{\mathrm{dof}} = n_{\mathrm{exp}} - n_{\mathrm{par}} \gg 1$, 
where $n_{\mathrm{exp}}$ and $n_{\mathrm{par}}$ are, respectively, 
the number of $R^{\mathrm{had}}_{ik}$ measurements 
and  the number of free parameters in the $R^{\mathrm{had}}(s)$ parameterization.
The $\chi^2$ value outside this range indicates either an inadequacy of the $R^{\mathrm{had}}(s)$ parameterization
or significant tensions between $R^{\mathrm{had}}$ values measured by different experiments.
The former possibility can be excluded by identifying subsets of experiments, 
such that any experiment belongs to at least one subset, all subsets cover a representative $s$ range, and any subset alone yields a consistent fit.
As this is always the case in our analysis, poor $\chi^2$ values obtained in the fits of 
$\pi^+\pi^-(\gamma)$ (with $\chi^2/n_{\mathrm{dof}} = 2.11$, $n_{\mathrm{dof}} = 754$)
and other channels using all available data \cite{ihep-cs} can be attributed to systematic tensions between certain experiments.

The commonly recommended recipe for such cases is to scale {\it all} experimental uncertainties, both statistical and systematic, 
by overall Birge factor $\sqrt{\chi^2/n_{\mathrm{dof}}}$ \cite{ParticleDataGroup:2026aaa}, 
the central values of $R^{\mathrm{had}}_{ik}$ (and, hence, $R^{\mathrm{had}}_{fit}(s)$) being unchanged.
However, in our case this recipe is inadequate for (at least) the following reason.
The statistical uncertainty is determined by an exactly known number of reconstructed signal events 
and a normalization factor depending on the luminosity, detector acceptance, reconstruction efficiency, etc.
Birge scaling  of statistical uncertainties in all experiments implies simultaneous scaling of their normalizations by the same factor, which was not intended. 
On the other hand, systematic uncertainties may be underestimated, or some sources of systematics may be overlooked at all.
Thus, in order to account for tensions between experiments, it is feasible to modify only their systematic uncertainties, keeping the statistical ones as is.

An additional systematic uncertainty accounting for the tensions can be found from the data as outlined below.
The core idea is to identify experiments with a maximally non-uniform distribution of contributions to $\chi^2$ over degrees of freedom 
as an indicator of the tension source, 
and estimate the additional systematic uncertainty from the discrepancies between these experiments and the fit.
In what follows, the $\pi^+\pi^-(\gamma)$ channel is used as an illustration. 

For our purposes, it is sufficient to project the discrepancy vector of each experiment,
\begin{equation} \label{eq:delta}
	\Delta_{ik} \equiv  R^{\mathrm{had}}_{ik} - \frac{1}{(\Delta \sqrt{s_{ik}})} \int\limits_{(\Delta \sqrt{s_{ik}})} R^{\mathrm{had}}_{fit}(s) d\sqrt{s} \, ,
\end{equation}
onto unit eigenvectors of the covariance matrix $C$
and find the projections yielding contributions $\Delta\chi^2 \gg 1$ to the $\chi^2$.
Let us define the systematic uncertainty vector for the individual experiment and normalize it to~1:
\begin{equation} \label{eq:vsys}
	v_{\mathrm{sys}, ik}  =  \Delta_{\mathrm{sys}, ik} \cdot \frac{1}{(\Delta\sqrt{s_{ik}})} \int\limits_{(\Delta\sqrt{s_{ik}})} R^{\mathrm{had}}_{fit}(s) d\sqrt{s}  \, , \,\, %
	 n_{\mathrm{sys}, ik}  = %
		   v_{\mathrm{sys}, ik} %
                  \left/  %
				  \sqrt{ \sum\limits_l v_{\mathrm{sys}, il}^2 } %
				  \right. %
		   \, .
\end{equation}
Taking $n_{\mathrm{sys}, ik}$ as an approximation for the eigenvector of the covariance matrix (\ref{eq:C})
corresponding to systematics of the $i$-th experiment,
project $\Delta_{ik}$ onto $n_{\mathrm{sys}, ik}$ and compute the projection's contribution to $\chi^2$ ($s$ bin indices are omitted):
\begin{equation} \label{eq:delta-sys}
	\Delta_{\mathrm{sys}, i} = n_{\mathrm{sys}, i} \left(n_{\mathrm{sys}, i}^T \Delta_i\right) \, , \,\, %
	\Delta\chi^2_{\mathrm{sys}, i} = \Delta_{\mathrm{sys}, i}^T C^{-1}_{i\, i} \Delta_{\mathrm{sys}, i} \, .
\end{equation}
The residual discrepancy (orthogonal to $\Delta_{\mathrm{sys}, i}$ and corresponding to point-to-point statistical fluctuations) 
and its contribution to $\chi^2$ are then:
\begin{equation}\label{eq:res}
	\Delta_{\mathrm{res}, i} = \Delta_i - \Delta_{\mathrm{sys}, i}\, , \,\, %
	\Delta\chi^2_{\mathrm{res}, i} = \Delta_{\mathrm{res}, i}^T C^{-1}_{i\, i} \Delta_{\mathrm{res}, i} \, .
\end{equation}
In a consistent fit,
one expects all experiments to have $\Delta\chi^2_{\mathrm{sys}, i} \sim 1$ and $\Delta\chi^2_{\mathrm{res}, i} \sim n_{p, i}$, 
where $n_{p, i}$ is the number of $s$ points measured in the $i$-th experiment.%
\footnote{%
In other terms, besides a good $\chi^2$, 
the individual contributions to the latter from all degrees of freedom should be distributed with a maximum entropy.
}
The actual values of $\Delta\chi^2_{\mathrm{sys}}$ and $\Delta\chi^2_{\mathrm{res}}/n_p$ for the experiments contributing to 
the $\pi^+\pi^-(\gamma)$ channel are shown in the Table~\ref{tab:pull-2pi} in the 3rd column.
The table also shows total contributions of the experiments to $\chi^2$ and their relative integral pulls with respect to the fit:
\begin{equation}\label{eq:pull}
	\mathrm{Pull}_i = \frac{
						\sum\limits_k   R^{\mathrm{had}}_{ik} \Delta \sqrt{s_{ik}} 
					  }{
					    \sum\limits_l \int\limits_{(\Delta\sqrt{s_{il}})} R^{\mathrm{had}}_{fit}(s) d\sqrt{s}
					  }
					  \, - \, 1
\end{equation}

\begin{table}[H] 
  \caption{%
	The $R^{\pi^+\pi^-(\gamma)}$ fits with unmodified and modified systematic uncertainties:
	$\chi^2$ contributions from individual experiments and from the systematic (\ref{eq:delta-sys})
	and residual (\ref{eq:res}) projections of the experiment {vs} fit discrepancies (\ref{eq:delta}).
	$n_p$ is the number of $s$ points contributing to the fit.
	The integral pull is defined by (\ref{eq:pull}).
	References to the data are given in \cite{ihep-cs}.
	The estimated extra systematic uncertainty is  $\epsilon = 5.7$\%.
  }
  \label{tab:pull-2pi}
  \begin{center}
	  {
	  \begin{tabular}{ll | r | rrr r | rrr r }
\hline
	Exp. & Run & $n_p$ & %
		 \multicolumn{4}{c|}{Unmodified systematics} & %
		 \multicolumn{4}{c}{Extra    systematics}   %
		  \\
	\multicolumn{2}{c|}{}	       & & %
          $\frac{\Delta\chi^2}{n_p}$ & $\Delta\chi^2_{sys}$ & $\frac{\Delta\chi^2_{res}}{n_p}$ & Pull & %
          $\frac{\Delta\chi^2}{n_p}$ & $\Delta\chi^2_{sys}$ & $\frac{\Delta\chi^2_{res}}{n_p}$ & Pull   %
		 \\[1ex]
\hline
  CMD-3  & 2020         &     13  & 4.08 & \bf 44.06 & 0.95  &   0.048 &    1.11 &  1.00 & 0.96 &  0.052 \\
         & 2018         &    114  & 2.00 & \bf 93.80 & 1.12  &   0.071 &    1.11 &  5.36 & 1.01 &  0.058 \\
         & 2013         &     82  & 1.79 & \bf 54.15 & 1.00  &   0.072 &    1.14 &  3.20 & 0.98 &  0.059 \\
  BaBar  & 2012         &    323  & 1.32 & \bf 13.45 & 1.23  &   0.020 &    1.19 &  0.05 & 1.18 &  0.007 \\
  KLOE   & 2018         &     85  & 2.97 &      1.94 &\bf3.00&  -0.009 &    2.14 &  1.39 & 2.27 & -0.022 \\
  BES-III& 2015         &     60  & 0.86 &      0.29 & 0.85  &   0.002 &    0.80 &  0.02 & 0.81 & -0.011 \\
  CLEO-c & 2018         &     35  & 0.67 &      1.36 & 0.63  &   0.016 &    0.66 &  0.02 & 0.65 &  0.003 \\
  CMD-2  & 2003         &     43  & 0.90 &      8.37 & 0.81  &   0.018 &    0.87 &  0.06 & 0.85 &  0.007 \\
         & 2005         &     36  & 0.75 &      0.46 & 0.66  &   0.071 &    0.74 &  0.12 & 0.68 &  0.037 \\
         & 2006         &     10  & 1.71 &      8.10 & 1.35  &  -0.052 &    1.21 &  1.10 & 1.29 & -0.056 \\
		 & 2006($\rho$) &     29  & 1.14 &      1.83 & 1.05  &   0.011 &    1.01 &  0.00 & 1.01 & -0.001 \\
  SND    & 2006         &     45  & 1.54 &      0.01 & 1.55  &  -0.001 &    1.37 &  0.08 & 1.42 & -0.012 \\
         & 2021         &     36  & 2.15 &      2.73 & 2.05  &   0.013 &    2.08 &  0.00 & 2.07 &  0.002 \\
  CMD    & 1985         &     24  & 1.56 &      0.14 & 1.58  &  -0.007 &    1.56 &  0.00 & 1.56 & -0.017 \\
  OLYA   & 1985         &     79  & 0.97 &      0.00 & 0.97  &   0.004 &    0.87 &  0.07 & 0.90 & -0.009 \\
  DM1    & 1978         &     16  & 0.82 &      1.49 & 0.75  &  -0.042 &    0.75 &  0.65 & 0.77 & -0.053 \\
\hline
\end{tabular}

	  }
\end{center}
\end{table}

\noindent
Table~\ref{tab:pull-2pi} shows that certain experiments have $\Delta\chi^2_{\mathrm{sys}} \gg 1$, 
with the residuals (\ref{eq:res}) yielding $\Delta\chi^2_{\mathrm{res}}/n_p \sim 1$, 
which reflects tensions seen in Fig.~\ref{fig:2pi}.\footnote{%
	$\Delta\chi^2_{\mathrm{res}}/n_p \simeq 3$ for KLOE-2 may indicate that its covariance matrix is inadequately parameterized in the form (\ref{eq:C}).
	The exact covariance matrix referenced in \cite{KLOE-2:2017fda} was published only online at now unavailable URL \cite{KLOE-covariance} and is not used here.
}

\begin{figure}[H]
	\begin{center}
		\includegraphics[width=0.73\textwidth,clip]{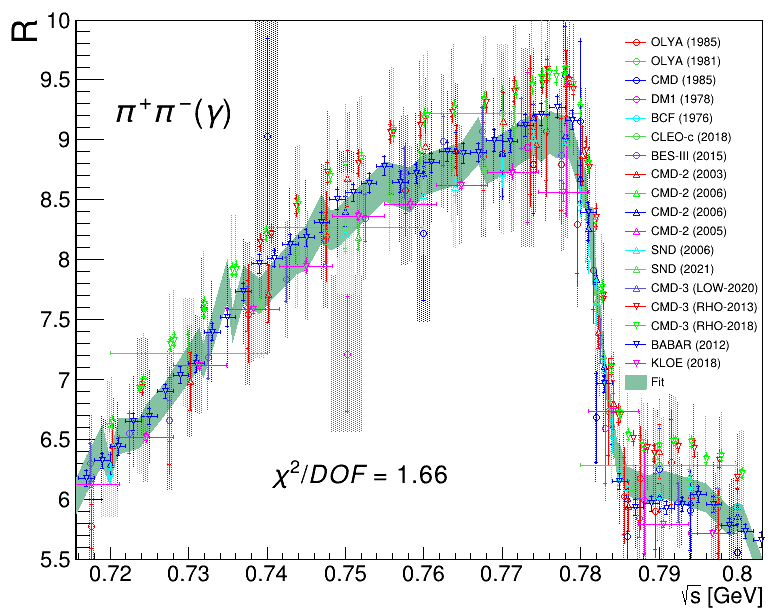}

		(a)
		
		\includegraphics[width=0.73\textwidth,clip]{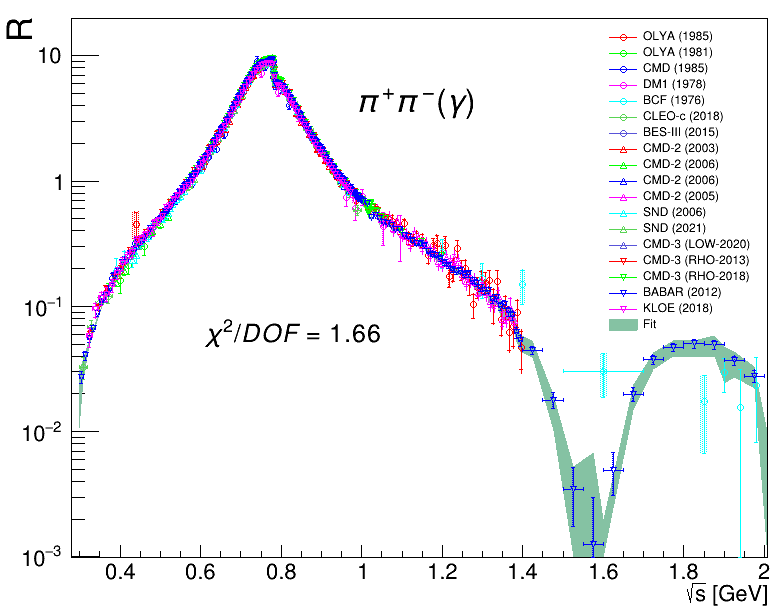}

		(b)
	\end{center}
	\vspace*{-2ex}
	\caption{(a) Same as in Fig.~\ref{fig:2pi}
			 after the  fit with the additional 5.7\% systematic uncertainty accounting for tensions between the experiments. 
			 The expanded total experimental uncertainties are shown by shaded rectangles.
			 The fit uncertainty (shown by green band) is scaled by $\sqrt{\chi^2/n_{\mathrm{dof}}}$ to account for residual tensions.
			 (b) The full $s$ range of the $\pi^+\pi^-(\gamma)$ fit.
	} \label{fig:2pi-mod}
\end{figure}

We estimate the additional uncertainty $\epsilon$ accounting for the tension as 
\begin{equation} \label{eq:xtra}
	\epsilon = \sqrt{ \langle \mathrm{Pull}_{\, i}^2 \rangle }\, , \,\, \Delta\chi^2_{\mathrm{sys}, i} > \chi^2_{thr}\, , 
\end{equation}
with the nominal threshold value $\chi^2_{thr} = 10$,
and add $\epsilon$ in quadrature as an extra normalization uncertainty to all experiments in the channel, 
assuming no correlation of pulls between them.
The modified covariance matrix thus reads:
\begin{equation} \label{eq:Cmod}
	C_{ik\, jl}^{\mathrm{mod}} = C_{ik\, jl} + \epsilon^2 \cdot \delta_{ij}  R^{\mathrm{had}}_{fit}(s_{ik})  R^{\mathrm{had}}_{fit}(s_{jl}) \, ,
\end{equation}
where $C$ is given by (\ref{eq:C}). 
The fit is then repeated  with the modified covariance matrix (\ref{eq:Cmod}),
and the individual contributions to $\chi^2$ are re-evaluated.
If all $\Delta\chi^2_{\mathrm{sys}, i} \le \chi^2_{thr}$, no further iterations are performed.\footnote{
There is no reason to iterate further 
as the adopted parameterization of the additional systematic uncertainty 
may be insufficient to completely exclude tensions and achieve $\chi^2/n_{\mathrm{dof}} \simeq 1$.}
Otherwise, the new $\epsilon$ value is computed according to  (\ref{eq:xtra}) 
and added to  $C^{\mathrm{mod}}$ in quadrature as in (\ref{eq:Cmod}).
In the $\pi^+\pi^-(\gamma)$ channel, the process converges after one iteration with the resulting $\epsilon = 5.7\%$ and 
all $\Delta\chi^2_{\mathrm{sys}}$ values below the threshold $\chi^2_{\mathrm{thr}} = 10$, as shown in Table~\ref{tab:pull-2pi} in the 4th column.
The result of the fit with the extra uncertainty $\epsilon$ is shown in Fig.~\ref{fig:2pi-mod}. 
The uncertainty of the fitted $R^{\pi^+\pi^-(\gamma)}(s)$ is of an order of discrepancy between BaBar, CMD-3 and KLOE-2 measurements,
far exceeding the underestimated Birge scaled uncertainty of the fit with the unmodified covariance matrices (Fig.~\ref{fig:2pi}).

\begin{table}[H] 
  \caption{%
	  Same as in Table~\ref{tab:pull-2pi} for the $R^{\pi^+\pi^-\pi^0}$ fits.
	  For the SND (2003) dataset the condition  $\Delta\chi^2_{\mathrm{sys}} < 10$ was not achieved in four iterations, 
	  with the estimated extra systematic uncertainty $\epsilon = 9.7$\%.
	  The process was aborted at this point as further expansion of $\epsilon$ 
	  resulted in growth of $\Delta\chi^2_{\mathrm{sys}}$. 
	  This indicates that the adopted approximation for the extra uncertainty is insufficient to account for tensions present in the 
	  $\pi^+\pi^-\pi^0$ channel.
	  References to the data are given in \cite{ihep-cs}.
	  CMD-2 (1995, 1998) datasets lack information on systematic uncertainties, 
	  hence $\Delta\chi^2_{\mathrm{sys}, \mathrm{res}}$ were not computed for them after the initial fit.
  }
  \label{tab:pull-3pi}
  \begin{center}
	  {
	  
\begin{tabular}{ll r | rrr r | rrr r }
\hline
	Exp. & Run & $n_p$ & %
		 \multicolumn{4}{c|}{Original systematics} & %
		 \multicolumn{4}{c}{Extra    systematics}   %
		  \\
		 &     &      &  %
          $\frac{\Delta\chi^2}{n_p}$ & $\Delta\chi^2_{sys}$ & $\frac{\Delta\chi^2_{res}}{n_p}$ & Pull & %
          $\frac{\Delta\chi^2}{n_p}$ & $\Delta\chi^2_{sys}$ & $\frac{\Delta\chi^2_{res}}{n_p}$ & Pull   %
		 \\[.5ex]
\hline
Belle & 2024 &  212 &  1.45 & {\bf 69.92}  & 1.23 &  0.048  &  1.09 &  2.95 & 0.98 &  0.025  \\
	SND   & 2003 &   67 &  2.37 & {\bf 150.81} & 1.48 &  0.073  &  1.32 & {\bf 15.21} & 1.17 &  0.058  \\
\hline
SND   & 2026 &  102 &  0.47 &  5.56 & 0.47 & -0.002  &  0.41 &  2.80 & 0.45 & -0.035  \\ 
SND   & 2000 &   32 &  2.09 &  3.06 & 1.64 &  0.071  &  1.74 &  0.43 & 1.58 &  0.056  \\
SND   & 2003 &   49 &  1.39 &  0.01 & 1.38 &  0.004  &  1.10 &  0.16 & 1.17 & -0.021  \\
SND   & 2015 &   40 &  1.13 &  1.48 & 0.99 &  0.071  &  0.90 &  0.48 & 0.79 &  0.067  \\
BaBar & 2005 &   75 &  1.04 &  4.39 & 0.80 &  0.120  &  0.92 &  1.76 & 0.73 &  0.105  \\
CMD-2 & 2000 &   13 &  2.37 &  5.52 & 2.32 & -0.028  &  2.76 &  8.53 & 2.48 & -0.050  \\
CMD-2 & 1998 &   16 &  2.08 &       &      &         &  1.33 &  2.82 & 0.94 &  0.050  \\
CMD-2 & 1995 &   14 &  1.13 &       &      &         &  0.93 &  1.24 & 0.81 &  0.026  \\
ND    & 1991 &   39 &  1.37 &  1.23 & 1.63 & -0.125  &  1.34 &  1.42 & 1.68 & -0.146  \\
CMD   & 1989 &    7 &  1.28 &  2.78 & 2.10 &  0.154  &  1.19 &  1.78 & 1.61 &  0.138  \\
DM1   & 1989 &   26 &  1.07 &  0.08 & 1.04 &  0.019  &  1.12 &  0.00 & 1.13 & -0.001  \\
\hline
\end{tabular}

	  }
\end{center}
\end{table}

\begin{figure}[H]
	\begin{center}
		\includegraphics[width=1.0\textwidth,clip]{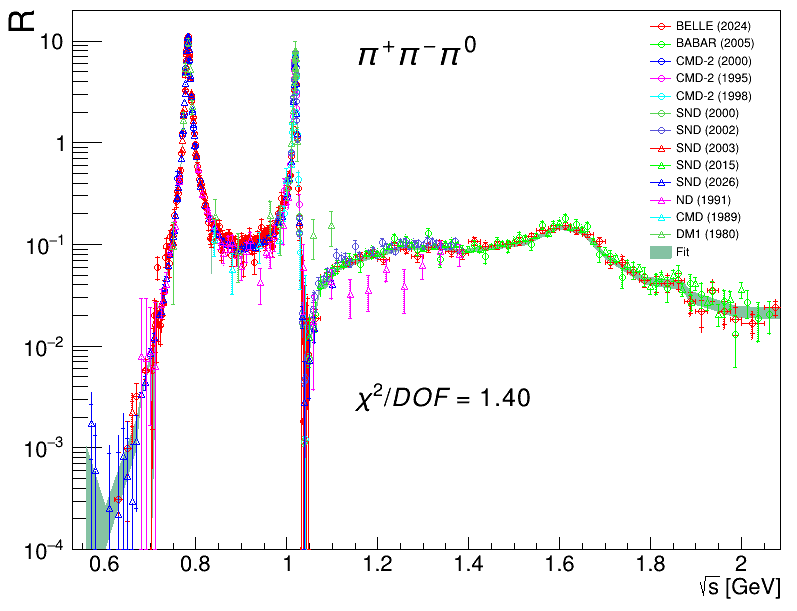}\\
		\vspace*{-3ex}
		(a)
		\vspace*{4ex}
		
		\includegraphics[width=0.48\textwidth,clip]{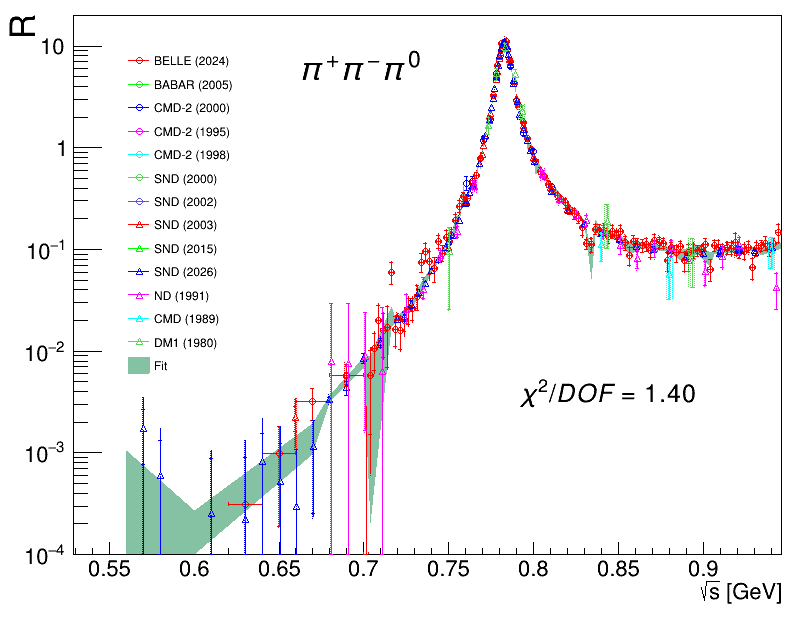}
		\includegraphics[width=0.48\textwidth,clip]{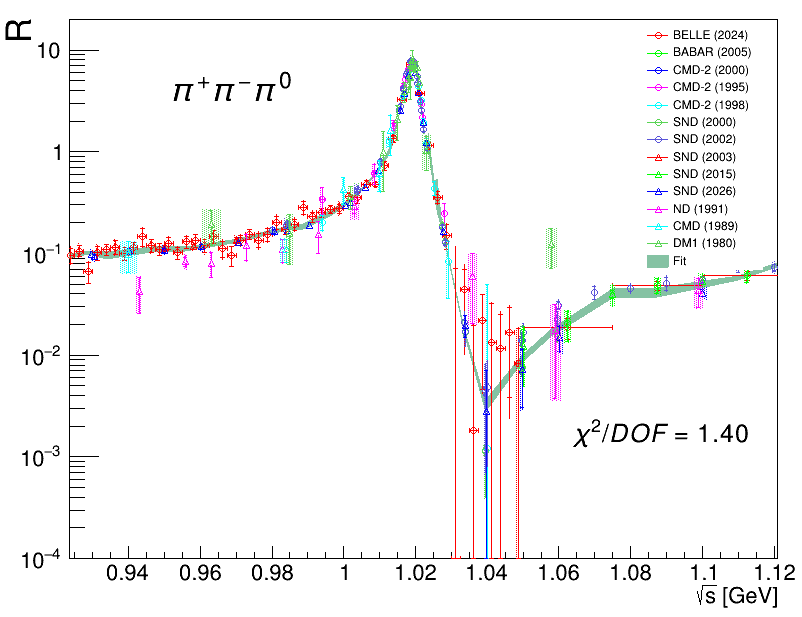}

		\hspace*{0.25\textwidth} (b) \hfill (c) \hspace*{0.2\textwidth}
	\end{center}
	\caption{%
		(a) The final fit of $R^{\pi^+\pi^-\pi^0}$ with $\epsilon = 9.7$\%. 
	    (b), (c) Enlarged view of $\omega(783)$ and $\phi(1020)$ regions.
	} \label{fig:3pi-mod}
\end{figure}

The above procedure is applied to all channels.
The extra systematic uncertainties $\epsilon$ are identified only for 
$\pi^+\pi^-(\gamma)$, $\pi^+\pi^-\pi^0$, $K^+K^-$, $\eta\gamma$ and $\omega\pi$.%
\footnote{Tensions are also present in the inclusive $R^{\mathrm{had}}$ measurements at $\sqrt{s} > 2$~GeV, 
though they fall below our threshold for the modification of systematics. 
See, e.g., the discussion of tensions in the  $R^{\mathrm{had}}(s)$ measurements 
below open charm production threshold in Ref.~\cite{Kataev:2026gea} and references therein.
}
The condition $\Delta\chi^2_{\mathrm{sys}, i} \le 10$ is satisfied upon the first iteration of covariance matrix modification (\ref{eq:Cmod})
in all these channels,  except $\pi^+\pi^-\pi^0$. 
In the latter, the process is aborted on fourth iteration where minimum values of $\Delta\chi^2_{\mathrm{sys}, i}$ are achieved (Table~\ref{tab:pull-3pi}).
Further iterations result in a growth of $\Delta\chi^2_{\mathrm{sys}, i}$, indicating that 
the adopted approximation for the extra uncertainty is insufficient to account for tensions present in the channel.
The result of the $R^{\pi^+\pi^-\pi^0}$ fit with the estimated $\epsilon = 9.7$\% is shown in Fig.~\ref{fig:3pi-mod}.

Apparently, the parameterization of the extra uncertainty in the form (\ref{eq:Cmod}) is insufficient to completely account for tensions 
and can be considered only as the first approximation. 
A more consistent procedure may involve determination of an $s$-dependent $\epsilon$, 
by simultaneosly looking for a minimum of $\chi^2$ and a maximum of the ``entropy'' of individual contributions to the $\chi^2$ from individual degrees of freedom.
Further studies are required to refine these criteria and implement the proper numerical procedure.

In absence of such a procedure,  residual tensions in the fits are accounted for by scaling 
the uncertainty of $R^{\mathrm{had}}_{fit}(s)$
by Birge factor $\sqrt{\chi^2/n_{\mathrm{dof}}}$, in case $P(\chi^2, n_{\mathrm{dof}}) < 0.05$.

\section{Results}

The contributions to $\ahad$ from individual channels are shown in Table~\ref{tab:channels}.
At $s$ below experimental $\pi^+\pi^-$ and $\pi^0\gamma$ production thresholds the ChPT form-factors are used.
The 3-loop pQCD expression for $R^{\mathrm{had}}(s)$ is used at $\sqrt{s} > 11.2$~GeV.
Contributions of narrow $\Psi(1S,2S)$ and $\Upsilon$ resonances are computed using their Breit--Wigner parameterization \cite{Bryzgalov:2024ebj}. 

The total LO hadronic contribution to $a_\mu$ is
\begin{equation}\label{eq:ahad}
	\ahad = \left( 697.7 \pm 9.8_{e^+e^-} \pm 1.1_{\chi^2_{thr}} \pm 2.3_{par} \pm 2.5_{rad} \right) \times 10^{-10}\, ,
\end{equation}
where the first uncertainty is due to $e^+e^- \to hadrons$ input data
(including the extra systematic uncertainty accounting for tensions between them),
the second one is related to variation of the $\chi^2_{thr}$ parameter (\ref{eq:xtra}) in the $6 < \chi^2_{thr} < 25$ interval,
the third is the systematic uncertainty of $R^{\mathrm{had}}(s)$ parameterization,
and the last one is due to radiative corrections.

\begin{figure}[H]
	\begin{center}
		\includegraphics[width=0.8\textwidth,clip]{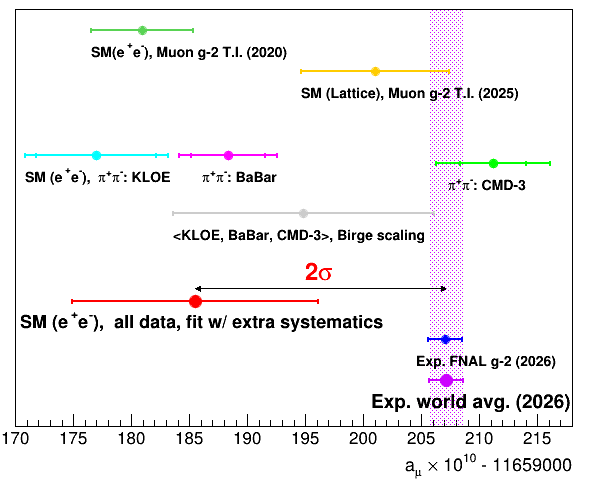}
	\end{center}
	\caption{
	The experimental world average $a_\mu^{\mathrm{exp}}$ 
	dominated by the FNAL g-2 measurement \cite{Muong-2:2026qnz}
	{vs} theoretical $a_\mu^{\mathrm{SM}}$ values including various $\ahad$ estimates (from top to bottom):
	the average of dispersion estimates \cite{Aoyama:2020ynm} before publication of CMD-3 $\pi^+\pi^-$ data;
	lattice QCD estimate~\cite{Muong-2:2026qnz};
	our dispersion estimates with the $\pi^+\pi^-(\gamma)$ contribution using only 
	KLOE-2 \cite{KLOE-2:2017fda}, BaBar \cite{BaBar:2012bdw} and CMD-3 \cite{CMD-3:2023alj} 
	(supplemented by OLYA and BCF data \cite{ihep-cs} to cover the region $1 < \sqrt{s} < 2$ GeV, no tension in the $\pi^+\pi^-(\gamma)$ fit),
	below the latter their weighed average with an uncertainty scaled by Birge factor is shown;
	our dispersion estimate (\ref{eq:ahad}) using $R^{\mathrm{had}}$ refitted with 
	additional systematic uncertainties $\epsilon$ accounting for tensions between 
	$\sigma_{\mathrm{tot}}(e^+e^- \to hadrons)$ measurements by different experiments.
	} \label{fig:status}
\end{figure}

\section {Conclusion}

Adding the $\ahad$ value (\ref{eq:ahad}) 
to the known electromagnetic, electroweak and higher order hadronic $a_\mu$ terms \cite{Aliberti:2025beg},
we obtain the SM value of the muon anomaly 
\begin{equation}
	a_\mu^{\mathrm{SM}} =  (11659185.5 \pm 10.6) \times 10^{-10}\, , \nonumber
\end{equation}
below the experimental world average $a_\mu^{\mathrm{exp}}$ \cite{Muong-2:2026qnz} at $2\sigma$ level.
The uncertainty is dominated by tensions between $e^+ e^- \to \pi^+\pi^-(\gamma)$ total cross sections 
measured in several experiments with 1\% level precision.
Summary of recent theoretical $a_\mu$ estimates compared to its experimental world average is shown in Fig.~\ref{fig:status}.

The precision of the SM $a_\mu$ estimate using the $e^+e^-$ based dispersive $\ahad$ is 
insufficient to claim a significant discrepancy between theoretical and experimental values of $a_\mu$.

New precise measurements of $\sigma_{\mathrm{tot}}(e^+e^- \to hadrons)$ are anticipated, 
via direct energy scan or radiative return at operating BEPCII, SuperKEKB, VEPP-2000, VEPP-4M, 
and future STCF \cite{STCF}, VEPP-6 \cite{VEPP-6} colliders.

\begin{table}[htbp] 
  \caption{%
	  Contributions to $\ahad$ from individual final states.
	  $\chi^2/n_{\mathrm{dof}}$ values are shown for the final fit with 
	  the extra systematic uncertainty $\epsilon$ accounting for tensions between the experiments.
	  The sources of uncertainties are:
	  (exp.) -- experimental uncertainty of the $e^+e^-$ input data 
				(including the extra uncertainty $\epsilon$ 
				 and scaled by $\sqrt{\chi^2/n_{\mathrm{dof}}}$ if $P(\chi^2, n_{\mathrm{dof}}) < 0.05$);
	  ($\chi^2_{thr}$) -- variation of the $\chi^2_{thr}$ threshold in (\ref{eq:xtra});
	  (par.) -- $R^{\mathrm{had}}(s)$ parameterization;
	  (rad.) -- radiative corrections.
  }
  \label{tab:channels}
  \begin{center}
	  \small
	  \begin{tabular}{lrrcrr}
	Final state & \parbox{0.25\textwidth}{$a_\mu(\mathrm{had},\mathrm{LO})$ $\times 10^{10}$\\ {(exp.) ($\chi^2_{\mathrm{thr}}$) (par.) (rad.)}}  & $\sqrt{s}\, [\mathrm{GeV}]$ & $\epsilon$, \% & $\frac{\chi^2}{\mathrm{dof}}$ & d.o.f. \\[0.5ex]
\hline
\hline
	$ \pi^+ \pi^- (\gamma) $ & 509.006 (9.631) (0.372) (2.280) (2.294) & 0.3 $ \div $ 1.937 & 5.7$^{+0.8}_{-0.5}$ & 1.66 & 754 \\
	$ \pi^+ \pi^- \pi^0 $ & 48.238 (1.785) (0.572) (0.201) (0.063) & 0.56024 $ \div $ 1.937 & 9.7$^{+0.0}_{-3.5}$ & 1.40 & 531 \\
$ \pi^+ \pi^- 2\pi^0 $ & 19.294 (0.436) (0.000) (0.063) (0.064) & 0.85 $ \div $ 1.937 &  & 1.17 & 76 \\
$ 2\pi^+ 2\pi^- $ & 14.411 (0.174) (0.000) (0.183) (0.012) & 0.6125 $ \div $ 1.93 &  & 1.41 & 124 \\
$ K^+ K^- $ & 23.198 (0.203) (0.447) (0.089) (0.008) & 0.985 $ \div $ 1.937 & 0.0$^{+6.6}_{-0.0}$  & 2.35 & 188 \\
$ K_S K_L $ & 13.106 (0.106) (0.000) (0.000) (0.000) & 1.00028 $ \div $ 1.937 &  & 0.95 & 157 \\
$ \pi^0 \gamma $ & 4.359 (0.093) (0.000) (0.049) (0.000) & 0.59986 $ \div $ 1.38 &  & 1.70 & 75 \\
$ K_S K^+ \pi^- + K_S K^- \pi^+ $ & 1.814 (0.100) (0.000) (0.000) (0.000) & 1.24 $ \div $ 1.937 &  & 0.99 & 20 \\
$ 2\pi^+ 2\pi^- \pi^0 $ & 1.219 (0.076) (0.000) (0.017) (0.001) & 1.0125 $ \div $ 1.937 &  & 0.58 & 29 \\
$ 2\pi^+ 2\pi^0 2\pi^- $ & 1.381 (0.141) (0.000) (0.011) (0.000) & 1.3125 $ \div $ 1.937 &  & 1.49 & 2 \\
$ 2\pi^+ 2\pi^- 3\pi^0 $ & 0.099 (0.013) (0.000) (0.002) (0.001) & 1.575 $ \div $ 1.937 &  & 0.57 & 1 \\
$ 3\pi^+ 3\pi^- $ & 0.254 (0.014) (0.000) (0.002) (0.012) & 1.3125 $ \div $ 1.937 &  & 1.52 & 54 \\
$ 3\pi^+ 3\pi^- \pi^0 $ & 0.020 (0.004) (0.000) (0.001) (0.000) & 1.6 $ \div $ 1.937 &  & 0.65 & 1 \\
$ \eta \gamma $ & 0.640 (0.062) (0.002) (0.047) (0.000) & 0.59986 $ \div $ 1.354 & 5.9$^{+0.0}_{-5.9}$ & 2.05 & 122 \\
$ \eta \pi^+ \pi^- $ & 0.575 (0.019) (0.000) (0.000) (0.000) & 1.15 $ \div $ 1.937 &  & 1.18 & 72 \\
$ K^+ K^- \pi^0 $ & 0.202 (0.050) (0.000) (0.000) (0.001) & 1.44 $ \div $ 1.937 &  & 0.54 & 16 \\
$ K^+ K^- \pi^0 \pi^0 $ & 0.100 (0.010) (0.000) (0.000) (0.000) & 1.5 $ \div $ 1.937 &  & 1.32 & 8 \\
$ K^+ K^- \pi^+ \pi^- $ & 0.810 (0.038) (0.000) (0.009) (0.000) & 1.4125 $ \div $ 1.937 &  & 1.91 & 38 \\
$ K^+ K^- \pi^+ \pi^- \pi^0 $ & 0.129 (0.019) (0.000) (0.000) (0.000) & 1.6125 $ \div $ 1.937 &  & 1.63 & 11 \\
$ K_S K_L \eta $ & 0.238 (0.052) (0.000) (0.000) (0.000) & 1.575 $ \div $ 1.937 &  & 1.31 & 6 \\
$ K_S K_L \pi^0 $ & 0.839 (0.093) (0.000) (0.000) (0.000) & 1.425 $ \div $ 1.937 &  & 1.50 & 6 \\
$ K_S K_L \pi^0 \pi^0 $ & 0.137 (0.043) (0.000) (0.000) (0.000) & 1.35 $ \div $ 1.937 &  & 0.00 & 0 \\
$ K_S K_L \pi^+ \pi^- $ & 0.166 (0.028) (0.000) (0.000) (0.000) & 1.425 $ \div $ 1.937 &  & 0.00 & 0 \\
$ K_S K^+ \pi^- \pi^0 + K_S K^- \pi^+ \pi^0 $ & 0.640 (0.044) (0.000) (0.000) (0.000) & 1.51 $ \div $ 1.937 &  & 1.08 & 15 \\
$ K_S K_S \pi^+ \pi^- $ & 0.066 (0.006) (0.000) (0.000) (0.000) & 1.63 $ \div $ 1.937 &  & 1.37 & 4 \\
$ \omega(783) \eta $ & 0.035 (0.002) (0.000) (0.000) (0.000) & 1.34 $ \div $ 1.937 &  & 0.85 & 42 \\
$ \omega(783) < \pi^0 \gamma > \pi^0 $ & 0.906 (0.022) (0.020) (0.102) (0.000) & 0.75 $ \div $ 1.937 & 0.0$^{+9.2}_{-0.0}$  & 1.56 & 135 \\
$ \omega(783) < \pi^+ \pi^- \pi^0 > \pi^+ \pi^- $ & 0.092 (0.005) (0.000) (0.000) (0.000) & 1.15 $ \div $ 1.937 &  & 0.90 & 67 \\
$ \omega \eta \pi^0 $ & 0.189 (0.045) (0.000) (0.091) (0.000) & 1.5 $ \div $ 1.937 &  & 0.30 & 8 \\
$ \pi^+ \pi^- 2\pi^0 \eta $ & 0.117 (0.019) (0.000) (0.000) (0.000) & 1.625 $ \div $ 1.937 &  & 0.85 & 5 \\
$ \pi^+ \pi^- 3\pi^0 $ & 1.067 (0.112) (0.000) (0.000) (0.000) & 1.125 $ \div $ 1.937 &  & 0.68 & 9 \\
$ \pi^+ \pi^- \pi^0 \eta $ & 0.663 (0.075) (0.000) (0.000) (0.000) & 1.394 $ \div $ 1.937 &  & 0.82 & 25 \\
$ \phi(1020) < X - K\bar{K} > \eta $ & 0.068 (0.003) (0.000) (0.001) (0.000) & 1.56 $ \div $ 1.937 &  & 0.97 & 74 \\
$ p \bar p $ & 0.033 (0.003) (0.000) (0.001) (0.000) & 1.889 $ \div $ 1.937 &  & 0.70 & 19 \\
$ n \bar n $ & 0.026 (0.004) (0.000) (0.000) (0.000) & 1.89 $ \div $ 1.937 &  & 1.48 & 8 \\
 2hadron (hadrons)  & 42.704 (0.552) (0.000) (0.235) (0.000) & 1.937 $ \div $ 11.199 &  & 1.39 & 314 \\
\hline
pQCD & 2.065 \phantom{(0.000)} (0.002) \phantom{(0.000)}   & $ > $ 11.1990  & & \\
ChPT $ \pi\pi, \pi^0\gamma $ & 0.538 \phantom{(0.000)} (0.013) \phantom{(0.000)}   &  0.2792  $ \div $ 0.3000  & & \\
\hline
$ \Psi(1S) $ &  6.495 \phantom{(0.000)} (0.124)  \phantom{(0.000)} &  3.0969 & & \\
$ \Psi(2S) $ &  1.631 \phantom{(0.000)} (0.057)  \phantom{(0.000)} &  3.6861 & & \\
$ \Upsilon(1S) $ &  0.054 \phantom{(0.000)} (0.002)  \phantom{(0.000)} &  9.4604 & & \\
$ \Upsilon(2S) $ &  0.021 \phantom{(0.000)} (0.003)  \phantom{(0.000)} & 10.0234 & & \\
$ \Upsilon(3S) $ &  0.014 \phantom{(0.000)} (0.002)  \phantom{(0.000)} & 10.3551 & & \\
$ \Upsilon(4S) $ &  0.010 \phantom{(0.000)} (0.001)  \phantom{(0.000)} & 10.5794 & & \\
\hline
\hline
 Total & {\bf697.671} (9.829) (1.103) (2.274) (2.455) &  &  & \\
\end{tabular}

\end{center}
\end{table}

\section*{Acknowledgments}
	The authors are grateful to V.~B.~Anikeev, A.~L.~Kataev, A.~G.~Myagkov
	and K.~Yu.~Todyshev for useful discussions.

\bibliographystyle{utphys}

\bibliography{biblio}

@article{ParticleDataGroup:2026aaa,
    author = "Takahashi, F. and others",
    collaboration = "Particle Data Group",
    title = "{Review of Particle Physics}",
    doi = "10.1142/S0217751X26300115",
    journal = "Int. J. Mod. Phys. A",
    volume = "41",
    pages = "2630011",
    year = "2026"
}

@article{Muong-2:2026qnz,
    author = "Aguillard, D. P. and others",
    collaboration = "Muon g{\ensuremath{-}}2",
    title = "{Final Report on the Measurement of the Positive Muon Anomalous Magnetic Moment at Fermilab to 127 ppb}",
    eprint = "2606.17323",
    archivePrefix = "arXiv",
    primaryClass = "hep-ex",
    reportNumber = "FERMILAB-PUB-26-0380-AD-PPD",
    month = "6",
    year = "2026"
}

@article{Aoyama:2020ynm,
    author = "Aoyama, T. and others",
    title = "{The anomalous magnetic moment of the muon in the Standard Model}",
    eprint = "2006.04822",
    archivePrefix = "arXiv",
    primaryClass = "hep-ph",
    reportNumber = "FERMILAB-PUB-20-207-T, INT-PUB-20-021, KEK Preprint 2020-5,
  MITP/20-028, KEK Preprint 2020-5, MITP/20-028, CERN-TH-2020-075, IFT-UAM/CSIC-20-74, LMU-ASC 18/20, LTH 1234,
  LU TP 20-20, LTH 1234, LU TP 20-20, MAN/HEP/2020/003, PSI-PR-20-06, UWThPh 2020-14, ZU-TH 18/20",
    doi = "10.1016/j.physrep.2020.07.006",
    journal = "Phys. Rept.",
    volume = "887",
    pages = "1--166",
    year = "2020"
}

@article{Aliberti:2025beg,
    author = "Aliberti, R. and others",
    title = "{The anomalous magnetic moment of the muon in the Standard Model: an update}",
    eprint = "2505.21476",
    archivePrefix = "arXiv",
    primaryClass = "hep-ph",
    reportNumber = "CERN-TH-2025-101, FERMILAB-PUB-25-0344-T, INT-PUB-25-015, IPARCOS-UCM-25-029, KEK Preprint 2025-22, LTH 1403, MITP-25-037, UWThPh 2025-15, UWThPh
  2025-15, ZU-TH 37/25, IPARCOS-UCM-25-029",
    doi = "10.1016/j.physrep.2025.08.002",
    journal = "Phys. Rept.",
    volume = "1143",
    pages = "1--158",
    year = "2025"
}

@article{Petermann:1957ir,
    author = "Petermann, A.",
    title = "{Magnetic moment of the mu meson}",
    reportNumber = "CERN-57-27",
    doi = "10.1103/PhysRev.105.1931",
    journal = "Phys. Rev.",
    volume = "105",
    pages = "1931",
    year = "1957"
}

@article{BaBar:2012bdw,
    author = "Lees, J. P. and others",
    collaboration = "BaBar",
    title = "{Precise Measurement of the $e^+ e^- \to \pi^+\pi^- (\gamma)$ Cross Section with the Initial-State Radiation Method at BABAR}",
    eprint = "1205.2228",
    archivePrefix = "arXiv",
    primaryClass = "hep-ex",
    reportNumber = "BABAR-PUB-12-003",
    doi = "10.1103/PhysRevD.86.032013",
    journal = "Phys. Rev. D",
    volume = "86",
    pages = "032013",
    year = "2012"
}

@article{CMD-3:2023alj,
    author = "Ignatov, F. V. and others",
    collaboration = "CMD-3",
    title = "{Measurement of the e+e-{\textrightarrow}{\ensuremath{\pi}}+{\ensuremath{\pi}}- cross section from threshold to 1.2~GeV with the CMD-3 detector}",
    eprint = "2302.08834",
    archivePrefix = "arXiv",
    primaryClass = "hep-ex",
    doi = "10.1103/PhysRevD.109.112002",
    journal = "Phys. Rev. D",
    volume = "109",
    number = "11",
    pages = "112002",
    year = "2024"
}

@article{KLOE-2:2017fda,
    author = "Anastasi, A. and others",
    collaboration = "KLOE-2",
    title = "{Combination of KLOE $\sigma\big(e^+e^-\rightarrow\pi^+\pi^-\gamma(\gamma)\big)$ measurements and determination of $a_{\mu}^{\pi^+\pi^-}$ in the energy range $0.10 < s < 0.95$ GeV$^2$}",
    eprint = "1711.03085",
    archivePrefix = "arXiv",
    primaryClass = "hep-ex",
    doi = "10.1007/JHEP03(2018)173",
    journal = "JHEP",
    volume = "03",
    pages = "173",
    year = "2018"
}

@article{Bryzgalov:2024ebj,
    author = "Bryzgalov, V. V. and Zenin, O. V.",
    title = "{Estimation of the LO Hadronic Contribution to g$_{\mu}$ {\textendash} 2 Using the NRC KI{\textemdash}IHEP Total Cross Section Database}",
    doi = "10.1134/S1063779624701053",
    journal = "Phys. Part. Nucl.",
    volume = "55",
    number = "6",
    pages = "1432--1438",
    year = "2024"
}

@misc{the-code,
    author = "Bryzgalov, V. V. and Zenin, O. V.",
	title = "{{\tt G-2HAD}: \rm the program for evaluation of $\ahad$ using $\sigma_{tot}(e^+e^- \to hadrons)$ data}",
	url = "https://glab.ihep.su/zenin_o/compas_users/-/tree/G-2HAD/rpp/ee/",
	language = "english"
}

@misc{ihep-cs,
	title = "{\rm NRC KI -- IHEP CrossSection database}",
	url = "http://hera.ihep.su:9991/ppds/bin/ee",
	language = "english"
}

@article{Kataev:2026gea,     
	author = "Kataev, A. L. and Todyshev, K. Yu.",     
	title = "{Perturbative QCD fitting of KEDR and BESIII $e^+e^-$ data for $R(s)$ and $\alpha_s$ determination}",     
	journal = "accepted to Natural Science Review", 
	volume = "4",
	number = "8",
	eprint = "2603.29803",     
	archivePrefix = "arXiv",     
	primaryClass = "hep-ph",     
	month = "3",     
	year = "2026" 
}

@misc{KLOE-covariance,
	collaboration = "KLOE-2",
	title = "{\rm KLOE combination (2017) $\pi^+\pi^-(\gamma)$ (ppg) data web link}",
	year = "2017",
	url ="https://www.lnf.infn.it/kloe/ppg/ppg_2017/ppg_2017.html"
}

@unpublished{STCF,
	collaboration = "STCF",
	title = "{\rm Progress of the Super Tau Charm Facility project in China}",
	author = "X. Qin and H. Peng and W. Yan",
	url = "https://indico.sns.it/event/140/contributions/1252/",
	year = "2026",
	note = "{14th Int. Workshop on $e^+e^-$ collisions from Phi to Psi 2026, 8-11 June 2026, Pisa, Italy}"
}

@unpublished{VEPP-6,
	title = "{\rm The VEPP-6 $e^+e^-$ collider project}",
	author = "Logashenko, I. B.",
	url = "https://indico.ihep.su/event/922/contributions/990/",
	year = "2026",
	note = "{Particle physics at intermediate and high energies, 2-5 June 2025, Protvino, Russia}"
}

\end{document}